\newif\ifshowchecklist
 \showchecklistfalse

\documentclass{article}

    \PassOptionsToPackage{numbers, compress}{natbib}

    \usepackage[preprint]{neurips_2025}

\usepackage[utf8]{inputenc} 
\usepackage[T1]{fontenc}    
\usepackage{hyperref}       
\usepackage{url}            
\usepackage{booktabs}       
\usepackage{amsfonts}       
\usepackage{nicefrac}       
\usepackage{microtype}      
\usepackage{xcolor}         
\usepackage{amsmath}
\usepackage[ruled]{algorithm2e}
\usepackage{amsthm}
\usepackage{multirow}

\usepackage{graphicx}

\newtheorem{theorem}{Theorem}
\newtheorem{condition}{Condition}
\newtheorem{property}{Property}

\newtheorem{proposition}{Proposition}

\newcommand{\pr}{\mathbb{P}}
\newcommand{\E}{\mathbb{E}}
\newcommand{\Cov}{\textnormal{Cov}}
\newcommand{\Var}{\textnormal{Var}}
\newcommand{\VCR}{\textnormal{V}_{\textnormal{CR}}}

\title{Fast Rerandomization via the BRAIN}

\author{%
  Jiuyao Lu \\
  University of Pennsylvania \\
  \texttt{jiuyaolu@wharton.upenn.edu} \\
  \And
  Daogao Liu \\
  University of Washington \\
  \texttt{dgliu@uw.edu} 
  \AND
  Zhanran Lin \\
  University of Pennsylvania \\
  \texttt{zhanranl@wharton.upenn.edu} \\
  \And
  Xiaomeng Wang \\
  University of Pennsylvania \\
  \texttt{xwang1@wharton.upenn.edu}
}

\begin{document}

\maketitle

\begin{abstract}
Randomized experiments are a crucial tool for causal inference in many different fields. Rerandomization addresses any covariate imbalance in such experiments by resampling treatment assignments until certain balance criteria are satisfied. However, rerandomization based on na\"ive acceptance-rejection sampling is computationally inefficient, especially when numerous independent assignments are required to perform randomization-based statistical inference. Existing acceleration methods are suboptimal and not applicable in structured experiments, including stratified and clustered experiments. Based on metaheuristics in integer programming, we propose BRAIN --- a novel computationally-lightweight methodology that ensures covariate balance in randomized experiments while significantly accelerating the computation. Our BRAIN method provides unbiased treatment effect estimators with reduced variance compared to complete randomization, preserving the desirable statistical properties of traditional rerandomization. Simulation studies and a real data example demonstrate the benefits of our method in fast sampling while retaining the appealing statistical guarantees.
 
\end{abstract}

\section{Introduction} \label{sec:intro}
Randomized experiments have been the gold standard for drawing causal inferences in a wide range of settings, including clinical trials \citep{rosenberger2015randomization}, social sciences \citep{bloom2008core}, and economic research \citep{duflo2007using}. 
The method relies on the assumption that, by randomly assigning units to treatment and control groups, covariates are balanced. Therefore, any differences in outcomes between the groups can be attributed causally to the treatment.
However, covariate imbalances between treatment groups often exist in practice due to the randomness in such complete randomization (CR) \citep{rosenberger2008handling,xu2010propensity,morgan2012rerandomization,krieger2019nearly}. 
This is a major issue because it introduces confounding, making it unclear whether observed differences in outcomes are due to the treatment or to pre-existing differences between groups, consequently invalidating any inference made.

An intuitive solution to this issue is rerandomization -- randomizing repeatedly until some prespecified covariate balance criterion is achieved \citep{student1938comparison,cox1982randomization,bailey1987valid,maclure2006measuring,imai2008misunderstandings,bruhn2009pursuit}. The most popular choice for the balance criterion is requiring the Mahalanobis distance between the covariate matrices of the treatment group and the control group to be lower than some threshold $a$. Despite the long history and widespread use of rerandomization, its theoretical implications using the Mahalanobis distance was only recently studied by \cite{morgan2012rerandomization}. Since then, rerandomization has attracted growing research interests, and its theory underpinnings have been established in various scenarios \citep{morgan2015tiers,li2018asymptotic,zhou2018sequential,li2020factorial,shi2022split,wang2023stratified,lu2023cluster,zhao2024no,branson2024power}.

 Classical rerandomization methods repeatedly sample treatment assignments until a balanced one is generated. However, in randomization-based inference, we need to generate a large number of balanced treatment assignments to ensure reliable estimation and inference. This often requires performing thousands of rerandomizations to obtain sufficiently many independent balanced assignments. As a consequence, rerandomization is extremely time-consuming using the acceptance-rejection sampling rerandomization (ARSRR) strategy \citep{luo2021leveraging,chung2018rapid,bind2020possible}. 
 This pose practical challenges, especially in large-scale or real-time experiments where treatment assignments must be generated quickly. High computational cost may limit the number of balanced assignments available for randomization-based inference, and thus discourage practitioners from adopting rerandomization methods altogether.

Several methods have been proposed to alleviate the computational burden of rerandomization. However, these approaches often fall short in practice, particularly in challenging settings involving diminishing covariate imbalance or high-dimensional covariates \citep{branson2021ridge,wang2022diminishing,zhang2023pca}. Meanwhile, some alternative approaches --- such as deterministic or semi-deterministic algorithms --- may yield a single acceptable assignment more quickly. 
Nevertheless, these methods typically lack the ability to generate independent assignments and do not ensure the unbiasedness of the resulting treatment effect estimator \citep{efron1971forcing,rosenberger2015randomization,johansson2021optimal}.

To address these limitations, \cite{zhu2022pair} introduced the Pair-Switching Rerandomization (PSRR) algorithm to accelerate the search for balanced assignments in both nonsequential and sequential experiments.
 The algorithm starts with a random assignment and gradually improves it by switching between a treated unit and a control unit repeatedly. Motivated by the Metropolis-Hastings algorithm \citep{metropolis1953equation,hastings1970monte}, PSRR allows moving to less balanced assignments with non-zero probabilities determined by a temperature hyperparameter to prevent being trapped in a local region. While the authors establish the unbiasedness and a lower bound on the variance reduction of the treatment effect estimator under PSRR, the PSRR method remains computationally intensive in many scenarios as shown in Section \ref{sec:simulation}.

Structured experiments are also common in practice. The ability to adapt to these special structures are also important for rerandomization algorithms. Rerandomization methods have been proposed for sequential experiments where units arrive group by group \citep{zhou2018sequential, bertsimas2019covariate}; stratified experiment where experimenters divide units into strata and conduct randomization within each stratum \citep{fisher1926arrangement, wang2023stratified}; and clustered experiments where all units in the same cluster receives the same treatment \citep{lu2023cluster}. The applicability of PSRR is limited to sequential experiments only. Beside CR and the na\"ive ARSRR, there does not exist any fast and unified framework that works in all these structured settings. Therefore, we aim to propose a method that is fast and adaptable to all structured settings. 


In this paper, we novelly consider reformulating the problem from a constrained optimization (rather than sampling) perspective. For an experiment with $n$ units, among which $n_t$ units are assigned to the treatment group and $n_c = n - n_t$ units are assigned to the control group; we denote $W = (W_1,\ldots,W_n)^T$ as the vector of treatment assignment indicators, where $W_i=1$ if unit $i$ is assigned to the treatment group and $W_i=0$ otherwise.
For some covariate balance criterion $M(W)$ --- for instance --- the popular choice of Mahalanobis distance \cite{morgan2012rerandomization},
we consider the optimization problem
\begin{align}\label{eq:bqp_intro}
    \textbf{minimize} &\quad M(W), \notag\\
    \textbf{subject to}  &\quad W \in \Big\{W \in \{0,1\}^n: \sum\nolimits_i W_i = n_t\Big\}.
\end{align}

Under this formulation, we consider utilizing combinatorial optimization techniques to conduct rerandomization with early termination. 
There are few studies concerning this specific optimization problem, but some related constrained binary quadratic programming problems have received extensive attention, such as the maximum diversity problem and the heaviest k-subgraph problem. These problems share the same feasible sets as \eqref{eq:bqp_intro} but have different objective functions. We recommend interested readers refer to \cite{marti2013heuristics} for a detailed review of the literature. 
The authors presented extensive experiments to reveal that the best methods to solve the maximum diversity problem are local search based metaheuristics, especially variable neighborhood search \citep{brimberg2009vns,aringhieri2011local} and tabu search \citep{palubeckis2007tabu,aringhieri2011local}. Both methods update their solutions by exchanging coordinates of $W$ that are ones with coordinates of $W$ that are zeroes. Variable neighborhood searching systematically changes the neighborhood by executing two steps iteratively. It descents to find local optimum in local search steps and perturbs to avoid being trapped in local regions in shaking steps. Tabu search maintains a list of forbidden moves, namely the tabu list, to avoid looping over recently visited coordinates of $W$.
Furthermore, \cite{aringhieri2011local} proposed to incorporate tabu search in the local search step of variable neighborhood search.
 

Following these intuitions, we propose a novel method to accelerate rerandomization --- Balanced RAdomization via INteger programming (BRAIN).
We show that a treatment assignment with a sufficiently small Mahalanobis distance can be achieved by optimizing a quadratic function of linearly constrained binary variables. Moreover, we prove that BRAIN, as a member of the rerandomization (RR) family, leads to treatment effect estimators that are unbiased and has lower variances than CR. Our simulation studies and real data analysis illustrate that BRAIN significantly improves the computational efficiency of rerandomization while maintaining the desired statistical properties. Using our method, the time it takes to obtain thousands of balanced treatment assignments can be reduced to seconds when existing methods require hours.

\section{Problem Setup} \label{sec:related}

In this section, we formally formulate the rerandomization framework and discuss existing approaches to solve the problem.

\textbf{The Neyman-Rubin Framework.}
We adopt the Neyman-Rubin potential outcome framework \citep{neyman1923application,rubin1974estimating}. We consider an experiment with $n$ units, among which $n_t$ units are assigned to the treatment group and $n_c = n - n_t$ units are assigned to the control group. We denote $W = (W_1,\ldots,W_n)^T$ as the vector of treatment assignment indicators, where $W_i=1$ if unit $i$ is assigned to the treatment group and $W_i=0$ otherwise. We assume that each unit has $p$-dimensional covariates, denoted by $X_i = (X_{i1},\ldots,X_{ip})^T$. We thus denote the covariate matrix as $X = (X_1,\ldots,X_n)^T$, with corresponding covariance matrix $S_{XX} = \frac{1}{n-1}\sum_{i=1}^n(X_i-\overline X)(X_i-\overline X)^T$. 
We assume that
each unit $i$ has two potential outcomes, $(Y_i(1),Y_i(0))$, which correspond to the two treatment arms. 

We denote the individual treatment effect for unit $i$ as $\tau_i = Y_i(1) - Y_i(0)$, and the average treatment effect  as $\tau = \frac{1}{n}\sum_{i=1}^n (Y_i(1) - Y_i(0))$. The observed outcome for unit $i$ is $Y_i = W_iY_i(1) + (1-W_i)Y_i(0)$. We adopt the stable unit treatment value assumption (SUTVA) \citep{rubin1980randomization}, which states that the treatment levels are well-defined, and there is no interference between the potential outcomes $(Y_i(1),Y_i(0))$ and the treatment assignments of other units $\{W_j: j \neq i\}$. We estimate the treatment effects with the difference-in-means estimator, defined as 
\[\widehat\tau(W) = \frac{1}{n_t}\sum_{i:W_i=1}Y_i(1) - \frac{1}{n_c}\sum_{i:W_i=0}Y_i(0).
\]

We employ randomization-based inference to construct confidence intervals for $\tau$ and test for the sharp null hypothesis $H_0: \tau_i=0$, $i = 1,\ldots,n$. See Appendix \ref{app:inference} for an introduction to randomization-based inference, including the Fisher randomization test. The procedure requires generating sufficiently many (e.g., 1000) i.i.d. assignments.


\textbf{Rerandomization with the Mahalanobis Distance.}
\cite{morgan2012rerandomization} suggested to conduct rerandomization by sampling $W$ from $\{W: \sum_{i=1}^n W_i = n_t, W_i \in \{0,1\}\}$ repeatedly until $M(W) \le a$, where $a$ is a prespecified threshold and $M(W)$ is the Mahalanobis distance between the covariates matrices of the treatment groups. $M(W)$ is formally defined as 
\begin{align*}
    M(W) &:= \left(\overline X_t - \overline X_c\right)^T \left[\Cov[\left(\overline X_t - \overline X_c\right)]\right]^{-1} \left(\overline X_t - \overline X_c\right) \\
    &= \left( W-\frac{n_t}{n}1_n \right)^T H \left( W-\frac{n_t}{n}1_n \right),
\end{align*}
where $H = \frac{n}{n_tn_c} X S_{XX}^{-1} X^T$.

Though this approach is commonly used in practice, it is computationally expensive. We will now introduce our BRAIN method, which addresses the computational challenge by leveraging metaheuristics from integer programming. 


\section{The BRAIN Method}\label{sec:BRAIN}
 In this section, we first introduce our Balanced RAdomization via INteger programming (BRAIN) algorithm for rerandomization. We then establish statistical properties of the resulting difference-in-means estimator using BRAIN.
\subsection{The BRAIN Algorithm}
We notice that rerandomization is closely related with the following constrained binary quadratic programming problem.
\begin{align}\label{eq:bqp}
    \textbf{minimize} &\quad M(W) = \left( W-\frac{n_t}{n}1_n \right)^T H \left( W-\frac{n_t}{n}1_n \right), \notag\\
    \textbf{subject to} &\quad \sum_{i=1}^n W_i = n_t, \notag\\
    &\quad W_i \in \{0,1\}, \; i=1,\ldots,n.
\end{align}

For simplification, we denote the feasible set of \eqref{eq:bqp} as $\mathcal{F}_0 = \{W \in \{0,1\}^n: \sum_i W_i = n_t\}$. While rerandomization aims to find a feasible $W$ in $\mathcal{F}_0$ with sufficiently small $M(W)$ rather than the minimizer of $M(W)$ in $\mathcal{F}_0$, we can still  utilize combinatorial optimization techniques to conduct rerandomization with early termination (as long as $M(W)$ falls below a prespecified threshold $a$). 

Motivated by the empirical success of local-search-based metaheuristics in related optimization problems, we propose the BRAIN method as follows. BRAIN starts with a random treatment assignment. Then, it conducts a \textbf{local search step}  where we sequentially examine $L$ disjoint random pairs of treated units and control units and swaps a pair if it leads to a more balanced assignment. If any swap has been made in the local search step, BRAIN will continue to search in the current local region. When no pairs are switched in the local search step, indicating a likely local optimum, BRAIN performs a \textbf{perturbation step} by randomly switching $S$ pairs of treated and control units. The algorithm will repeat the local search step and the perturbation step until reaching a satisfiable sample with $M(W)\leq \alpha$. The full algorithm is shown in Algorithm \ref{alg:BRAIN}.

BRAIN conducts the local search step differently from any existing variable neighborhood searching methods. We borrow the idea of preventing looping over recently visited units from tabu search and examine $L$ disjoint pairs. However, BRAIN differs from tabu search in the sense that BRAIN avoids maintaining the tabu list and checking whether each swap is forbidden, further improving computational efficiency. 

$L$ and $S$ are two hyperparameters in BRAIN. If $L$ is too small, the algorithm would still loop over recently visited pairs. If $S$ is too large, the perturbation is too strong such that it is similar to random restart. We compare different choices of $L$ and $S$ in Appendix \ref{app:results} and suggest setting $(L,S) = (n/2,1)$.

Our method differs from PSRR in that it accepts swaps only if $M(W)$ decreases in local search step. This way, it avoids inefficiently moving to a treatment assignment $W$ with higher $M(W)$, which frequently happens for PSRR. By avoiding this issue, our algorithm leads to significant faster descent of $M(W)$ and therefore much shorter computation time. PSRR with zero temperature is equivalent to BRAIN with $L=1$ and without the perturbation steps. Our experiments show that this special version of PSRR is faster than its default version but still much slower than BRAIN. 



\begin{algorithm}[htbp]
\caption{Balanced RAdomization via INteger programming (BRAIN)}
\label{alg:BRAIN}
\KwIn{Covariates $X$, group sizes $(n_t,n_c)$, threshold $a$, hyperparameter $(L,S)$.}
Set $t=0$, $W^{(0)}$ randomly with $n_t$ many 1's and $n_c$ many 0's, $M^{(0)} = M(W^{(0)})$\; 
\While{$M^{(t)} > a$}{
    Randomly select $L$ positions of the 1's in $W^{(t)}$, denoted by $O_1,\cdots,O_L$\;
    Randomly select $L$ positions of the 0's in $W^{(t)}$, denoted by $Z_1,\cdots,Z_L$\;
    Set $\widetilde S = S$\;
    \For{$l = 1, \ldots, L$}{
        Obtain $W^*$ by switching the 1 at position $O_l$ and the 0 at position $Z_l$ in $W^{(t)}$\;
        Set $M^* = M(W^*)$\;
        \If{$M^* < M^{(t)}$}{
            Set $\widetilde S = 0$, $t = t + 1$, $W^{(t)} = W^*$, $M^{(t)} = M^*$\;
            \lIf{$M^{(t)} \le a$}{break}
        }
    }
    \If{$\widetilde S > 0$}{
        Randomly select $\widetilde S$ positions of the 1's in $W^{(t)}$, denoted by $O_1,\cdots,O_S$\;
        Randomly select $\widetilde S$ positions of the 0's in $W^{(t)}$, denoted by $Z_1,\cdots,Z_S$\;
        \For{$s = 1, \ldots, \widetilde S$}{
            Obtain $W^{(t+1)}$ by switching the 1 at position $O_s$ and the 0 at position $Z_s$ in $W^{(t)}$\;
            Set $M^{(t+1)} = M(W^{(t+1)})$\;
            Set $t = t + 1$\;
        }
    }
}
\KwOut{$W = W^{(t)}$.}
\end{algorithm}
\vspace{10pt} 

\subsection{Statistical Properties of BRAIN}
In this section, we establish the statistical properties of BRAIN. We consider the following conditions, which are standard in the literature \citep{morgan2012rerandomization,zhu2022pair}.
\begin{condition} \label{cond:linear relationship} (linear decomposition)
    For $w_i = 0, 1$, $Y_i(w_i) = \beta_0 +\beta^TX_i +\tau w_i +e_i$, where $\beta_0 + \beta^TX_i$ is the linear projection of $Y_i(0)$ onto $(1,X)$ and $e_i$ is the deviation from the linear projection.
\end{condition}
\begin{condition} \label{cond:normality} (normality)
    $\widehat\tau$ and $\overline X_t - \overline X_c$ are normally distributed.
\end{condition}
\begin{condition} \label{cond:equal sizes} (equal group sizes)
    $n_t = n_c = \frac{n}{2}$.
\end{condition}

We prove that the difference-in-means estimator is unbiased under BRAIN.
\begin{theorem} \label{thm:unbiasedness}
    Under Condition \ref{cond:equal sizes}, if $W$ is generated by BRAIN, then $\E[\widehat\tau] = \tau$.
\end{theorem}

We also prove that the difference-in-means estimator has smaller variance than that under complete randomization and construct a lower bound for the variance reduction. Denote $\VCR$ as the variance of the difference-in-means estimator under complete randomization, we prove the following theorem.
\begin{theorem} \label{thm:variance}
    Under Conditions \ref{cond:linear relationship}-\ref{cond:equal sizes}, if $X$ has $p$ covariates and $W$ is generated by BRAIN with threshold $a$, then
\begin{align*}
    \frac{\VCR-\Var[\widehat\tau]}{\VCR} \ge \left( 1-\frac{a}{p} \right) R^2,
\end{align*}
where $$R^2 = \frac{\beta^T \Cov_{\textnormal{CR}}[\overline X_t - \overline X_c] \beta}{\VCR} = \frac{n\beta^T S_{XX} \beta}{n_tn_c\VCR}.$$
\end{theorem}
Here $\Cov_{\textnormal{CR}}[\overline X_t - \overline X_c]$ represents the covariance matrix of $\overline X_t - \overline X_c$ under complete randomization.

Like other rerandomization method, the difference-in-mean estimator still satisfies the same unbiasedness and variance reduction properties. This means that BRAIN is able to accelerate the process of rerandomization without sacrificing the desirable statistical properties.

\section{BRAIN for Structured Randomized Experiments} \label{sec:extension}

In this section, we extend BRAIN to randomized experiments with various types of structures -- sequential, stratified, and cluster randomized experiments. We will discuss how to adapt our algorithm to each of the three settings. The full algorithms are provided in Appendix \ref{app:pseudocode}, as well as the validation of their unbiasedness and variance reduction properties.

\subsection{BRAIN for Sequentially Randomized Experiments} \label{sec:SeqBRAIN}

In Section \ref{sec:BRAIN}, we consider offline rerandomization, which assigns treatment to all units at once. However, in sequential experimenters where participants arrive one by one or in groups, it is common to assign treatment to units group by group in practice. 

We assume that the experiment consists of $K$ stages such that the $n$ units are divided into $K$ sequential groups of sizes $(n_1,\ldots,n_K)$. In the $k$-th stage, $n_{t_k}$ units in the $k$-th group are assigned to the treatment arm, and the remaining $n_{c_k}$ units are assigned to the control arm. We use the subscript $[k]$ to represent the reduced version of any quantity calculated based on only the first $k$ groups. For example, $M_{[k]}(W_{[k]})$ represents the Mahalanobis distance of the covariate matrices of the first $k$ group. Sequential rerandomization aims to do the following: when the k-th group enrolls in the experiment, randomly assign treatment to the units without changing the assignment status of the previously enrolled groups, while balancing the covariates of all units in the first k groups with $M_{[k]}(W_{[k]}) \le a_k$.


We extend our BRAIN method to sequentially randomized experiments (denoted by SeqBRAIN, details in Algorithm \ref{alg:SeqBRAIN}) where it performs BRAIN in each stage of the experiment while fixing the assignments obtained in previous stages.





\subsection{BRAIN for Stratified Randomized Experiments} \label{sec:StratBRAIN}

Other than rerandomization, stratification is a common alternative approach to balance covariates. In stratified randomized experiments, experimenters divide units into strata, namely, layers or levels, according to the discrete covariates and conduct complete randomization within each stratum. If stratum-specific covariate balance is of interest, we can directly apply BRAIN to each strata. If the overall covariate balance is of interest, we can modify BRAIN to effectively account for this structure. 

We assume that the $n$ units are divided into $K$ strata of sizes $(n_1,\ldots,n_K)$. The set of indices for units in the $k$-th stratum is denoted by $\mathcal{I}_k$. In the $k$-th stratum, $n_{t_k}$ units are assigned to the treatment arm and $n_{c_k} = n_k-n_{t_k}$ units are assigned to the control arm. Similar to Section \ref{sec:related}, the assignment status is represented by a $n$-dimensional vector $W$ and the covariates are gathered into a $n \times p$ matrix $X$. The 'overall' strategy proposed by \cite{wang2023stratified} samples assignments from the space $\mathcal{W}_\textnormal{strat} = \{W \in \{0,1\}^n: \sum_{i \in \mathcal{I}_k} W_i = n_{t_k}, \forall k \in [K]\}$ repeatedly until $M(W) \le a$. To our knowledge, the only existing method to achieve this goal is to perform acceptance-rejection sampling, which we refer to as StratARSRR. 

We note that rerandomization in stratified experiments is equivalent to minimizing $M(W)$ under the constraint that $W\in \mathcal{W}_\textnormal{strat}$. We can extend BRAIN to stratified randomized experiments and refer to this variant as StratBRAIN (Algorithm \ref{alg:StratBRAIN}). StratBRAIN performs local searching and perturbation iteratively while considering the constraints imposed by the stratum structure. In the local search step, StratBRAIN randomly selects a list of multiple disjoint pairs of treated units and control units from each stratum. Then StratBRAIN aggregates the lists from different strata and permutes the pairs in the aggregated list. StratBRAIN follows this new ordering of unit pairs to examine whether switching each pair leads to a more balanced assignment. Compared to consecutively examining pairs in the same stratum, this procedure avoids any predetermined ordering of units or strata and allows for more flexible search in the local region. In the perturbation step, StratBRAIN conducts random unit swaps within every stratum. This ensures that the new assignment is sufficiently distant from the current one.


\subsection{BRAIN for Cluster Randomized Experiments} \label{sec:ClustBRAIN}
Many public health and social science experiments \citep{donner2000design,hayes2017cluster,schochet2020analyzing} assign the treatment at the cluster level rather than the individual level due to practical limitations or policy factors. Suppose the $n$ units are divided into $K$ clusters of sizes $(n_1,\ldots,n_K)$ in a cluster randomized experiment. We denote $\mathcal{I}_k$ as the index set of units in the $k$-th cluster. Cluster randomization assigns $K_t$ clusters to the treatment arm and $K_c = K - K_t$ clusters to the control arm. We note that when the $n_i$'s are not all equal, neither the treatment arm nor the control arm has a fixed number of units, which is different from all the settings we consider in previous sections. The assignment status of the clusters is denoted by a $K$-dimensional vector, $U$. And the assignment status of all units is denoted by a $n$-dimensional vector, $W$. $W_i=U_k$ if the $k$-th cluster includes the $i$-th unit.

\cite{lu2023cluster} suggested conducting cluster rerandomization to balance covariates in cluster randomized experiments. The authors proposed two metrics that measure the covariate imbalance. One is the Mahalanobis distance of cluster-level covariates, and the other is the Mahalanobis distance of individual-level covariates. Since using the first metric is equal to treating a cluster as an aggregated unit and can be efficiently solved by the approach proposed in Section \ref{sec:BRAIN}, we focus on how to modify BRAIN to minimize the second metric. 
We thus propose the ClustBRAIN method (Algorithm \ref{alg:ClustBRAIN}), a variant of BRAIN, which performs local searching and perturbation at the cluster level while treating the clusters as the operational units.

The detailed algorithms for each of the three cases are deferred to Appendix \ref{app:pseudocode}. In all three structured settings, the difference-in-means estimator resulting from our proposed methods enjoys similar unbiasedness and variance reduction properties to those in Section \ref{sec:BRAIN}. The details and proofs are shown in Appendix \ref{app:proofs}.

\section{Simulation Studies and Real Applications} \label{sec:simulation}

This section compares BRAIN with CR, ARSRR, and PSRR. To compare the computation time of BRAIN and the existing methods, we use each method to sample 1000 acceptable assignments respectively in one simulation and replicate the simulation 100 times on a MacBook Pro (Apple M1 Max Chip, 32 GB Memory, 10 Cores). To examine the statistical properties (such as standard deviation of the treatment effect estimator and length of confidence interval), we use each method to sample one acceptable assignment in the design stage to conduct the experiment and 1000 acceptable assignments in the analysis stage to conduct randomization-based inference. We replicate the simulation 1000 times using a High-Performance Computing cluster.

We note that CR is expected to be the fastest method with the worst statistical performances as it has no requirements for covariate balance at all. ARSRR, PSRR, and BRAIN are expected to have similar performances in terms of the statistical properties because all of them belong to the rerandomization family and use the same criterion $M(W) \le a$. Therefore, the ideal performance of BRAIN would be:
\begin{enumerate}
    \item Statistical properties: similar to ARSRR and PSRR, better than CR;
    \item Computational time: (much) faster than ARSRR and PSRR, but not CR.
\end{enumerate}

\subsection{Simple Randomized Experiments} \label{sec:simulation simple}

We first consider the simple randomized experiments described in Section \ref{sec:BRAIN} and mimic the simulation settings in \cite{zhu2022pair}. We assume that the covariates follow the standard normal distribution identically and independently, $X_{ij} \overset{i.i.d.}{\sim} N(0,1)$, $i = 1,\ldots,n$, $j = 1,\ldots,p$. The potential outcomes of the control group are generated independently from a linear model, $Y_i(0) = \sum_{j=1}^p X_{ij} + \epsilon_i$, where $\epsilon_i \sim N(0,p)$. We conduct simulations under both the null hypothesis and the alternative hypothesis:
\begin{align*}
& H_0: Y_i(1) = Y_i(0), \\ & H_a: Y_i(1) = Y_i(0) + 0.2\sqrt{\Var[Y_i(0)]}.
\end{align*}
 We consider three sample sizes, $n \in \{30,100,500\}$. The size of the treatment group and the control group are set as equal.

\cite{morgan2012rerandomization} proved that the asymptotic acceptance probability of the assignment is $p_a = P(\chi_p^2 \le a)$ and \cite{li2018asymptotic} recommended setting $a$ such that $p_a = 10^{-3}$. \cite{wang2022diminishing} considered settings where $p_a$ and $p$ are allowed to grow with $n$ rather than fixed. Motivated by these works, 
we consider two acceptance probabilities, $p_a \in \{10^{-3}, 10^{-4}\}$, and calculate the threshold $a$ by solving $p_a = P(\chi_p^2 \le a)$. We consider two regimes for the number of covariates, $p=2$ and $p=n/2$.

We perform Fisher randomization tests with a significance level $\alpha = 0.1$ and construct randomization-based confidence intervals with a nominal coverage rate of 90\%. 
For each method, we examine the absolute bias and standard deviation of the difference-in-means estimator, the size and power of the Fisher randomization test, the coverage and length of confidence intervals, and the average run time to sample 1000 acceptable assignments.
We also use the randomness metric in \cite{zhu2022pair} to measure the randomness of different methods. Recall that we need 1000 i.i.d. assignments to conduct randomization-based inference. If a method is almost deterministic or lacks randomness, its statistical inference performance can be impaired.

\begin{table}[htbp]
\caption{Statistical and computational performance of different methods in simple randomized experiments. SD: standard deviation. CP: coverage probability. Len: length.}

\label{tab:simple}
\begin{center}
\begin{tabular}{ccclccccccccc}
\toprule
         \multirow{2}{*}{$n$} & \multirow{2}{*}{$p$} & \multirow{2}{*}{$p_a$} & \multirow{2}{*}{Method} & & \multicolumn{6}{c}{Statistical Inference ($\times 10^{-3}$)} & & \multirow{2}{1.7cm}{\parbox{1.7cm}{\vspace{1ex}\centering Run Time \\ \centering (second)}} \\
         \cmidrule{6-11}
          & & & & & Bias & SD & Size & Power & CP & Len & & \\
          \hline
         \multirow{4}{*}{30} & \multirow{4}{*}{2} & \multirow{4}{*}{$10^{-3}$} & CR & & 6.2 & 74 & 10.5 & 14 & 89.5 & 246 & & $<0.1$ \\
         & & & ARSRR & & 0.1 & 51 & 8.7 & 19 & 91.3 & 177 & & 12.7 \\
         & & & PSRR & & 1.5 & 51 & 10.1 & 18 & 89.9 & 177 & & 50.3 \\
         & & & BRAIN & & 1.6 & 52 & 11.0 & 17 & 89.0 & 177 & & \textbf{1.4} \\
         \hline
         \multirow{3}{*}{500} & \multirow{3}{*}{250} & \multirow{3}{*}{$10^{-4}$} & CR & & 0.8 & 201 & 10.4 & 88 & 89.6 & 660 & & $<0.1$ \\
         & & & PSRR & & 4.3 & 187 & 10.3 & 91 & 89.7 & 611 & & 2878.8 \\
         & & & BRAIN & & 5.7 & 181 & 9.8 & 92 & 90.2 & 608 & & \textbf{1.3} \\
    \bottomrule
\end{tabular}
\end{center}
\end{table}

Table \ref{tab:simple} shows the statistical and computational performances of BRAIN and the other methods in two typical settings: a low-dimensional setting with $(n,p,p_a)=(30,2,10^{-3})$ and a high-dimensional setting with $(n,p,p_a)=(500,250,10^{-4})$. ARSRR is omitted from the high-dimensional setting since it is too slow to be a realistic option in this setting. The complete results (including ARSRR in moderately challenging settings) are provided in the  Appendix \ref{app:results}. From the results, we can see that BRAIN is significantly faster than ARSRR and PSRR in all settings. BRAIN merely requires seconds to generate 1000 acceptable assignments, even in cases where ARSRR or PSRR requires an hour. In addition, the computation time of BRAIN does not scale with the sample size $n$, covariate size $p$, or the threshold $p_a$ significantly as the other methods, making it an ideal algorithm to use, especially in high-dimensional settings. 

For all methods, the finite-sample biases are negligible, Fisher randomization test controls the type-1 error, and the confidence intervals are valid. Compared to CR, all the rerandomization methods reduce the variance of the difference-in-means estimator, increase the statistical power, produce tighter confidence intervals, and have comparable performances in terms of these aspects. BRAIN generates assignments that are as random as those generated by ARSRR, allowing us to conduct reliable statistical inference (see Appendix \ref{app:results}). These results demonstrate that BRAIN maintains the appealing statistical properties of rerandomization and substantially reduces the computational cost in simple randomized experiments.

\subsection{Structured Randomized Experiments} \label{sec:simulation structured}
This section considers the three types of structured experiments described in Section \ref{sec:extension}. We simulate the covariates and potential outcomes in the same way as in Section \ref{sec:simulation simple}, except that we set $Y_i(1) = Y_i(0) + 0.15\sqrt{\Var[Y_i(0)]}$ under the alternative to ensure the power of different methods remains within a moderate range. For sequential experiments, we set the number of groups as $K=2$ and consider two group sizes, $n_k \in \{100, 500\}$. We use the strategy proposed by \cite{zhou2018sequential} to determine $a_k$. Specifically, we set $(s_1,s_2) = (239, 761)$ when $p = 50$ and $(s_1,s_2) = (264, 736)$ when $p=250$, where $a_k=\frac{n_k}{n_{[k]}}q_k$ and $q_k$ is the $\frac{1}{s_k}$ quantile of a non-central chi-square distribution with $p$ degrees of freedom and a non-central parameter $\frac{n_{[k-1]}}{n_k}M_{[k-1]}$. For stratified experiments, we set the stratum number as $K=2$ and consider two stratum sizes, $n_k \in \{100, 500\}$. For cluster randomized experiments, we set the cluster size as $n_k=2$ and consider two options for the number of clusters, $K \in \{100,500\}$. We set $p_a=P(\chi_p^2 \le a)=10^{-3}$ for both stratified randomized experiments and cluster randomized experiments and set $p=n/4$ for all settings in this section.

Table \ref{tab:structured} shows the results when $n=1000$. The complete results are provided in Appendix \ref{app:results}. For clarity, we use a unified name to refer to different method variants. Since PSRR is not extended to stratified experiments and cluster experiments, it is excluded from the two schemes. The overall findings are consistent with the results in Section \ref{sec:simulation simple}. BRAIN substantially reduces the computational burden of rerandomization, especially for high-dimensional datasets and in stratified randomized experiments and cluster randomized experiments, where ARSRR is the only method available to perform rerandomization. BRAIN maintains the statistical properties of rerandomization in all the settings and significantly improves statistical inference compared to CR.

\begin{table}[htbp]
\caption{Statistical and computational performance of different methods in structured randomized experiments. SD: standard deviation. CP: coverage probability. Len: length.} \label{tab:structured}
\begin{center}
\begin{tabular}{clccccccccc}
\toprule
         \multirow{2}{*}{Scheme} & \multirow{2}{*}{Method} & & \multicolumn{6}{c}{Statistical Inference ($\times 10^{-3}$)} & & \multirow{2}{1.7cm}{\parbox{1.7cm}{\vspace{1ex}\centering Run Time \\ \centering (second)}} \\
         \cmidrule{4-9}
          & & & Bias & SD & Size & Power & CP & Len & & \\
          \hline
         \multirow{4}{*}{Sequential} 
         & CR & & 3.2 & 146 & 10.7 & 76 & 89.3 & 466 & & $<0.1$ \\
         & ARSRR & & 6.0 & 139 & 12.2 & 80 & 87.8 & 428 & & 3288.5 \\
         & PSRR & & 3.3 & 135 & 11.5 & 82 & 88.5 & 429 & & 339.1 \\
         & BRAIN & & 5.8 & 138 & 11.4 & 81 & 88.6 & 428 & & \textbf{5.8} \\
         \hline
         \multirow{3}{*}{Stratified} 
         & CR & & 3.0 & 140 & 10.0 & 76 & 90.0 & 466 & & $<0.1$ \\
         & ARSRR & & 2.7 & 131 & 9.1 & 82 & 90.9 & 434 & & 3386.8 \\
         & BRAIN & & 0.0 & 130 & 10.5 & 83 & 89.5 & 436 & & \textbf{3.0} \\
         \hline
         \multirow{3}{*}{Cluster} 
         & CR & & 0.9 & 140 & 9.8 & 77 & 90.2 & 465 & & $<0.1$ \\
         & ARSRR & & 6.2 & 132 & 9.3 & 80 & 90.7 & 435 & & 3116.2 \\
         & BRAIN & & 3.3 & 131 & 10.1 & 82 & 89.9 & 436 & & \textbf{3.9} \\
    \bottomrule
\end{tabular}
\end{center}
\end{table}

\section{A Clinical Trial Example} \label{sec:real data}
In addition to the simulation studies, we analyze a real dataset from a phase 1 clinical trial \cite{austin2009balance} to illustrate the advantages of BRAIN. The goal of this trial was to evaluate the interactions between intravenous methamphetamine and oral reserpine. In this trial, 20 participants were randomly assigned to the treatment group, and 10 participants were assigned to the control group. We aim to balance three covariates: age, weight, and pre-treatment heart rate, which are significantly unbalanced in the original study, to better infer the treatment effects on post-treatment heart rate.

We set $a$ as the 0.001 quantile of $\chi_3^2$ and generate 10,000 acceptable assignments using each method. ARSRR takes approximately 2.3 minutes to complete the task, and PSRR takes more than half an hour. In contrast, BRAIN requires only 12 seconds, which demonstrates that BRAIN accelerates the rerandomization process significantly. Based on the original dataset, we simulate semi-synthetic data and examine the statistical performance of each method in Appendix \ref{app:results}. The results are consistent with our simulation studies, where BRAIN preserves the desired statistical properties of rerandomization.

\section{Conclusion} \label{sec:conclusion}
In this work, we propose a novel BRAIN method to perform rerandomization efficiently in various types of randomized experiments. We derive the unbiasedness and a lower bound for the variance reduction of the difference-in-means estimator under BRAIN. Our empirical results demonstrate that BRAIN maintains the appealing statistical properties of rerandomization and significantly reduces the computational cost compared to existing methods.

\textbf{Limitations} This work does not address randomized experiments with tiers of covariates. Rerandomization for such experiments requires searching for assignments that simultaneously balance the covariates within each tier. In future work, we plan to extend our method to this setting using multi-objective optimization techniques. Another limitation is that it is unclear whether other integer programming techniques (e.g., branch and bound) can be adapted to develop better rerandomization algorithms. We leave this as our future research direction.


{
\small

\bibliographystyle{plainnat}
\bibliography{refs}



}


\newpage
\appendix

\section*{Appendix}

In Appendix \ref{app:pseudocode}, we provide the pseudocode of the three BRAIN variants. In Appendix \ref{app:proofs}, we state and prove all the theoretical results. In Appendix \ref{app:inference}, we introduce the Fisher randomization test and randomization-based confidence intervals. In Appendix \ref{app:results}, we show the simulation results not included in the main text.

\section{Algorithmic Description of BRAIN Variants} \label{app:pseudocode}
\subsection{Algorithmic Description of SeqBRAIN}
The pseudocode of SeqBRAIN is presented in Algorithm \ref{alg:SeqBRAIN}. As mentioned in the main text, SeqBRAIN performs BRAIN in each experiment stage sequentially while fixing the assignments obtained in previous stages.

\subsection{Algorithmic Description of StratBRAIN}
The pseudocode of StratBRAIN is presented in Algorithm \ref{alg:StratBRAIN}. StratBRAIN differs from BRAIN in that there are linear constraints within each stratum, $\sum_{i \in \mathcal{I}_k}W_i = n_{tk}$. Therefore, StratBRAIN only exchanges units in the same stratum to ensure that the aforementioned linear constraints are always satisfied. StratBRAIN does not loop over the strata sequentially. Instead, after selecting the pairs of units to examine from each stratum, StratBRAIN aggregates the lists of pairs and randomly permutes the combined list. In this way, we ensure that all strata are equally treated, allowing for flexible exploration of the neighborhood without a predetermined ordering of the strata or the units.

\subsection{Algorithmic Description of ClustBRAIN}
The pseudocode of ClustBRAIN is presented in Algorithm \ref{alg:ClustBRAIN}. ClustBRAIN treats each cluster as a whole and uses the clusters as the operational units in the search process. In other words, ClustBRAIN essentially conducts BRAIN at the cluster level rather than the unit level. Recall that we denote the assignment status of the clusters as a $K$-dimensional vector $U$ and all units' assignment status as a $n$-dimensional vector $W$. We have the following constraint imposed by the cluster structure of the experiment. $W_i=U_k$ if the $k$-th cluster includes the $i$-th unit. Since $W$ is determined once we generate $U$, we denote $W(U)$ as the mapping from $U$ to $W$ in Algorithm \ref{alg:ClustBRAIN}. 

\newpage
\begin{algorithm}[H]
\caption{Sequential BRAIN}
\label{alg:SeqBRAIN}
\KwIn{Covariates $X$, \hfill\break number of units to assign to the two arms in each stage $(n_{t1},\ldots,n_{tK})$, $(n_{c1},\ldots,n_{cK})$, \hfill\break thresholds $(a_1,\ldots,a_K)$, \hfill\break hyperparameter $(L_1,\ldots,L_K)$, $(S_1,\ldots,S_K)$.}
\For{$k = 1, \ldots, K$}{
Set $t=0$\;
Initialize $W_{[k]}^{(0)}$: keep the elements obtained in previous stages fixed in $W_{[k]}^{(0)}$ if $k \ge 2$ and set the undetermined elements in $W_{[k]}^{(0)}$ as $n_{tk}$ elements equal to 1 and $n_{ck}$ elements equal to 0 with random positions\;
Set $M_{[k]}^{(0)} = M_{[k]}(W_{[k]}^{(0)})$\; 
\While{$M_{[k]}^{(t)} > a_k$}{
    Randomly select $L_k$ positions of the 1's in $W_{[k]}^{(t)}$, denoted by $O_1,\cdots,O_{L_k}$\;
    Randomly select $L_k$ positions of the 0's in $W_{[k]}^{(t)}$, denoted by $Z_1,\cdots,Z_{L_k}$\;
    Set $\widetilde S = S_k$\;
    \For{$l = 1, \ldots, L_k$}{
        Obtain $W^*$ by switching the 1 at position $O_l$ and the 0 at position $Z_l$ in $W_{[k]}^{(t)}$\;
        Set $M^* = M(W^*)$\;
        \If{$M^* < M^{(t)}$}{
            Set $\widetilde S = 0$\;
            Set $t = t + 1$\;
            Set $W_{[k]}^{(t)} = W^*$\;
            Set $M_{[k]}^{(t)} = M^*$\;
            \lIf{$M_{[k]}^{(t)} \le a_k$}{break}
        }
    }
    \If{$\widetilde S > 0$}{
        Randomly select $\widetilde S$ positions of the 1's in $W_{[k]}^{(t)}$, denoted by $O_1,\cdots,O_{S_k}$\;
        Randomly select $\widetilde S$ positions of the 0's in $W_{[k]}^{(t)}$, denoted by $Z_1,\cdots,Z_{S_k}$\;
        \For{$s = 1, \ldots, \widetilde S$}{
            Obtain $W_{[k]}^{(t+1)}$ by the 1 at positon $O_s$ and the 0 at positon $Z_s$ in $W_{[k]}^{(t)}$\;
            Set $M_{[k]}^{(t+1)} = M_{[k]}(W_{[k]}^{(t+1)})$\;
            Set $t = t + 1$\;
        }
    }
}
}
\KwOut{$W = W^{(t)}$.}
\end{algorithm}

\newpage
\begin{algorithm}[H]
\caption{StratBRAIN}
\label{alg:StratBRAIN}
\KwIn{Covariates $X$, \hfill\break index sets of units in each stratum $(\mathcal{I}_1,\ldots,\mathcal{I}_K)$, \hfill\break number of units to assign to the two arms in each stratum $(n_{t1},\ldots,n_{tK})$, $(n_{c1},\ldots,n_{cK})$, \hfill\break threshold $a$, \hfill\break hyperparameter $(L_1,\ldots,L_K)$, $(S_1,\ldots,S_K)$, $L_{\textnormal{total}}=\sum_kL_k$, $S_{\textnormal{total}}=\sum_kS_k$.}
Set $t=0$\;
Set $W^{(0)}$ as $n_{tk}$ elements in $\mathcal{I}_k$ equal to 1 and $n_{ck}$ elements in $\mathcal{I}_k$ equal to 0 with random positions for all $k \in \{1,\ldots,K\}$\;
Set $M^{(0)} = M(W^{(0)})$\; 
\While{$M^{(t)} > a$}{
    \For{$k = 1, \ldots, K$}{
        Randomly select $L_k$ positions of the 1's in $W^{(t)}$ that belong to $\mathcal{I}_k$, denoted by $O_{k1},\cdots,O_{kL_k}$\;
        Randomly select $L_k$ positions of the 0's in $W^{(t)}$ that belong to $\mathcal{I}_k$, denoted by $Z_{k1},\cdots,Z_{kL_k}$\;
    }
    Aggregate $\cup_{1 \le k \le K, 1 \le l \le L_k}\{O_{kl}\}$ and denote the union as $(O_1,\ldots,O_{L_{\textnormal{total}}})$ with random ordering\;
    Aggregate $\cup_{1 \le k \le K, 1 \le l \le L_k}\{Z_{kl}\}$ and denote the union as $(Z_1,\ldots,Z_{L_{\textnormal{total}}})$. The ordering of $(Z_1,\ldots,Z_{L_{\textnormal{total}}})$ is consistent with that of $(O_1,\ldots,O_{L_{\textnormal{\textnormal{total}}}})$\;
    Set $\widetilde S = S_{\textnormal{total}}$\;
    \For{$l = 1, \ldots, L_{\textnormal{total}}$}{
        Obtain $W^*$ by switching the 1 at position $O_l$ and the 0 at position $Z_l$ in $W^{(t)}$\;
        Set $M^* = M(W^*)$\;
        \If{$M^* < M^{(t)}$}{
            Set $\widetilde S = 0$\;
            Set $t = t + 1$\;
            Set $W^{(t)} = W^*$\;
            Set $M^{(t)} = M^*$\;
            \lIf{$M^{(t)} \le a$}{break}
        }
    }
    \If{$\widetilde S > 0$}{
        \For{$k = 1, \ldots, K$}{
            Randomly select $S_k$ positions of the 1's in $W^{(t)}$ that are in $\mathcal{I}_k$, denoted by $O_{k1},\cdots,O_{kS_k}$\;
            Randomly select $S_k$ positions of the 0's in $W^{(t)}$ that are in $\mathcal{I}_k$, denoted by $Z_{k1},\cdots,Z_{kS_k}$\;
        }
        Aggregate $\cup_{1 \le k \le K, 1 \le l \le S_k}\{O_{kl}\}$ and denote the union as $(O_1,\ldots,O_{S_{\textnormal{total}}})$\;
        Aggregate $\cup_{1 \le k \le K, 1 \le l \le S_k}\{Z_{kl}\}$ and denote the union as $(Z_1,\ldots,Z_{S_{\textnormal{total}}})$\;
        \For{$s = 1, \ldots, S_{\textnormal{total}}$}{
            Obtain $W^{(t+1)}$ by switching the 1 at position $O_s$ and the 0 at position $Z_s$ in $W^{(t)}$\;
            Set $M^{(t+1)} = M(W^{(t+1)})$\;
            Set $t = t + 1$\;
        }
    }
}
\KwOut{$W = W^{(t)}$.}
\end{algorithm}

\newpage
\begin{algorithm}[H]
\caption{ClustBRAIN}
\label{alg:ClustBRAIN}
\KwIn{Covariates $X$, \hfill\break index sets of units in each cluster $(\mathcal{I}_1,\ldots,\mathcal{I}_K)$, \hfill\break number of clusters in the treatment arm and the control arm $(K_t,K_c)$, \hfill\break threshold $a$, \hfill\break hyperparameter $(L,S)$.}
Set $t=0$\;
Set $U^{(0)}$ as $K_t$ elements equal to 1 and $K_c$ elements equal to 0 with random positions\;
Set $M^{(0)} = M(W(U^{(0)}))$\; 
\While{$M^{(t)} > a$}{
    Randomly select $L$ positions of the 1's in $U^{(t)}$, denoted by $O_1,\cdots,O_L$\;
    Randomly select $L$ positions of the 0's in $U^{(t)}$, denoted by $Z_1,\cdots,Z_L$\;
    Set $\widetilde S = S$\;
    \For{$l = 1, \ldots, L$}{
        Obtain $U^*$ by switching the 1 at position $O_l$ and the 0 at position $Z_l$ in $U^{(t)}$\;
        Set $M^* = M(W(U^*))$\;
        \If{$M^* < M^{(t)}$}{
            Set $\widetilde S = 0$\;
            Set $t = t + 1$\;
            Set $W^{(t)} = W^*$\;
            Set $U^{(t)} = U^*$\;
            \lIf{$M^{(t)} \le a$}{break}
        }
    }
    \If{$\widetilde S > 0$}{
        Randomly select $\widetilde S$ positions of the 1's in $U^{(t)}$, denoted by $O_1,\cdots,O_S$\;
        Randomly select $\widetilde S$ positions of the 0's in $U^{(t)}$, denoted by $Z_1,\cdots,Z_S$\;
        \For{$s = 1, \ldots, \widetilde S$}{
            Obtain $U^{(t+1)}$ by switching the 1 at position $O_s$ and the 0 at position $Z_s$ in $U^{(t)}$\;
            Set $M^{(t+1)} = M(W(U^{(t+1)}))$\;
            Set $t = t + 1$\;
        }
    }
}
\KwOut{$W = W(U^{(t)})$.}
\end{algorithm}

\newpage
\section{Additional Theoretical Results and Proofs} \label{app:proofs}
BRAIN, StratBRAIN, and ClustBRAIN all assign treatment in an offline manner. In contrast, SeqBRAIN performs online allocation and needs to be handled differently. We will first establish the theoretical properties of BRAIN, StratBRAIN, and ClustBRAIN, and then obtain the results of SeqBRAIN. 

We have proposed the following conditions in the main text, which are related to the offline methods.
\setcounter{condition}{0}
\begin{condition} 
    For $w_i = 0, 1$, $Y_i(w_i) = \beta_0 +\beta^TX_i +\tau w_i +e_i$, where $\beta_0 + \beta^TX_i$ is the linear projection of $Y_i(0)$ onto $(1,X)$ and $e_i$ is the deviation from the linear projection.
\end{condition}

\begin{condition} 
    $\widehat\tau$ and $\overline X_t - \overline X_c$ are normally distributed.
\end{condition}

\begin{condition} 
    $n_t = n_c = \frac{n}{2}$.
\end{condition}

Similar to Condition \ref{cond:equal sizes}, we also consider the following equal-group-size conditions for stratified randomized experiments and cluster randomized experiments, respectively.

\begin{condition} \label{cond:equal sizes seq}
    $n_{t_k} = n_{c_k} = \frac{n_k}{2}$, $k = 1, \ldots, K$.
\end{condition}

\begin{condition} \label{cond:equal sizes clust}
    $K_{t} = K_{c} = \frac{K}{2}$.
\end{condition}




We point out three crucial properties of a rerandomization algorithm.

\begin{property} \label{prop:symmetric space} (symmetric feasible set)
    $\mathcal{F} = 1 - \mathcal{F}$.
\end{property}

\begin{property} \label{prop:uniform initialization} (uniform initialization)
    For any $w \in \mathcal{F}$,
    \begin{align*}
    \pr\left(W^{(0)} = w\right) = \frac{1}{|\mathcal{F}|}.
\end{align*}
\end{property}

\begin{property} \label{prop:symmetric update} (updating rule)
    There exists a sequence of functions $\{u_t\}_{t \ge 1}$, such that for any $t$ and $(w^{(0)},\ldots,w^{(t)})$ that may appear consecutively in the generated chain,
    \begin{align*}
        \pr\left(W^{(t)} = w^{(t)} \mid W^{(0)} = w^{(0)}, \ldots, W^{(t-1)} = w^{(t-1)}\right) = u_t(M(w^{(0)}),\ldots,M(w^{(t)})),
    \end{align*}
\end{property}

Any rerandomization algorithm with these properties can yield unbiased treatment effect estimators with reduced variances.
\begin{theorem} \label{thm:unbiasedness offline}
    If $W$ is generated by a rerandomization algorithm satisfying Properties \ref{prop:symmetric space}-\ref{prop:symmetric update}, then $E[\widehat\tau] = \tau$.
\end{theorem}

\begin{theorem} \label{thm:variance offline}
    Under Conditions \ref{cond:linear relationship} and \ref{cond:normality}, if $W$ is generated by a rerandomization algorithm satisfying Properties \ref{prop:symmetric space}-\ref{prop:symmetric update}, then
\begin{align*}
    \frac{\VCR-\Var[\widehat\tau]}{\VCR} \ge \left( 1-\frac{a}{p} \right) R^2,
\end{align*}
where $R^2 = \frac{n\beta^T S_{XX} \beta}{n_tn_c\VCR}$.
\end{theorem}

We shall verify that BRAIN, StratBRAIN, and ClustBRAIN satisfy these properties.
\setcounter{proposition}{0}
\begin{proposition} 
    BRAIN satisfies Properties \ref{prop:uniform initialization} and \ref{prop:symmetric update}. Under Condition \ref{cond:equal sizes}, BRAIN satisfies Property \ref{prop:symmetric space}.
\end{proposition}

\begin{proposition} 
    StratBRAIN satisfies Properties \ref{prop:uniform initialization} and \ref{prop:symmetric update}. Under Condition \ref{cond:equal sizes seq}, StratBRAIN satisfies Property \ref{prop:symmetric space}.
\end{proposition}

\begin{proposition} 
    ClustBRAIN satisfies Properties \ref{prop:uniform initialization} and \ref{prop:symmetric update}. Under Condition \ref{cond:equal sizes clust}, ClustBRAIN satisfies Property \ref{prop:symmetric space}.
\end{proposition}

Consequently, their unbiasedness and variance reduction results (including Theorems \ref{thm:unbiasedness} and \ref{thm:variance}) are immediately evident as corollaries of Theorems \ref{thm:unbiasedness offline} and \ref{thm:variance offline}.

In fact, the statistical guarantees of CR, ARSRR, and PSRR can also be established by verifying that they satisfy Properties \ref{prop:symmetric space}-\ref{prop:symmetric update}.




We will establish the statistical properties of SeqBRAIN by extending the conditions and properties in the offline setting to the online setting.

\subsection{Proof of Theorem \ref{thm:unbiasedness offline}}
For any chain $\left(w^{(0)}, \ldots, w^{(T)}\right)$ that the rerandomization algorithm may generate, we have 
\begin{align*}
    \pr\left(W^{(0)} = w^{(0)}\right) = \pr\left(W^{(0)} = 1-w^{(0)}\right),
\end{align*}
since the initial point $W(0)$ is sampled uniformly from a symmetric space $\mathcal{F}$.

If we exchange the assignments to the two arms by replacing $W$ with $1-W$, the covariate imbalance between the two arms should remain unchanged. In other words, $M(W) = M(1-W)$. As a result,
\begin{align*}
    &\hspace{15pt} \pr\left(W^{(t)} = w^{(t)} \mid W^{(0)} = w^{(0)}, \ldots, W^{(t-1)} = w^{(t-1)}\right) \\
    &= u_t\left( M(w^{(0)}),\ldots,M(w^{(t)}) \right) \\
    &= u_t\left( M(1-w^{(0)}),\ldots,M(1-w^{(t)}) \right) \\ 
    &= \pr\left(W^{(t)} = 1-w^{(t)} \mid W^{(0)} = 1-w^{(0)}, \ldots, W^{(t-1)} = 1-w^{(t-1)}\right),
\end{align*}
Therefore, for any $w $, we have
\begin{align*}
&\hspace{15pt} \pr\left(W^{(0)} = w^{(0)}, \ldots, W^{(T-1)} = w^{(T-1)}, W^{(T)} = w^{(T)}\right) \\
&= \pr\left(W^{(0)} = w^{(0)}\right) \pr\left(W^{(1)} = w^{(1)} \mid W^{(0)} = w^{(0)}\right) \cdots \pr\left(W^{(T)} = w^{(T)} \mid W^{(0)} = w^{(0)}, \ldots, W^{(T-1)} = w^{(T-1)}\right) \\
&= \pr\left(W^{(0)} = 1-w^{(0)}\right) \pr\left(W^{(1)} = 1-w^{(1)} \mid W^{(0)} = 1-w^{(0)}\right) \cdots \\
&\hspace{15pt} \pr\left(W^{(T)} = 1-w^{(T)} \mid W^{(0)} = 1-w^{(0)}, \ldots, W^{(T-1)} = 1-w^{(T-1)}\right) \\
&= \pr\left(W^{(0)} = 1-w^{(0)}, \ldots, W^{(T-1)} = 1-w^{(T-1)}, W^{(T)} = 1-w^{(T)}\right),
\end{align*}
We then sum over all possible $\left(w^{(0)}, \ldots, w^{(T-1)}\right)$ and obtain $\pr(W = w^{(T)}) = \pr(W = 1-w^{(T)})$. As a result, $\pr\left(W_i = 1\right) = \pr\left(W_i = 0\right) = 1 / 2$. And we have
\begin{align*}
\E[\widehat\tau(W)] &  = \E\left[\frac{2}{n} \sum_{i = 1}^n W_i Y_i(1)-\frac{2}{n} \sum_{i = 1}^n\left(1-W_i\right) Y_i(0)\right] \\
&= \frac{2}{n} \sum_{i = 1}^n Y_i(1) \pr\left(W_i = 1\right)-\frac{2}{n} \sum_{i = 1}^n Y_i(0) \pr\left(W_i = 0\right) \\
&= \tau .
\end{align*}

\subsection{Proof of Theorem \ref{thm:variance offline}}
The difference-in-means estimator can be expressed as
\begin{align*}
    \widehat{\tau} &= \tau+\beta^T\left(\overline X_t-\overline X_c\right)+\left(\overline e_t-\overline e_c\right).
\end{align*}
$\overline X_t - \overline X_c$ and $\overline e_t - \overline e_c$ are uncorrelated since $\beta_0+\beta^{\mathrm{T}} X_i$ is the projection of $Y_i(0)$ onto $(1,X)$. In addition, $\widehat\tau$ and $\overline X_t - \overline X_c$ are normally distributed, so $\overline X_t - \overline X_c$ and $\overline e_t - \overline e_c$ are independent. Since the rerandomization algorithm only affects $\overline X_t - \overline X_c$ and does not affect $\overline e_t - \overline e_c$, we have
\begin{align*}
\Var[\widehat\tau] &= \Cov\left[ \beta^T (\overline X_t-\overline X_c) \beta \right] + \Var\left[\overline e_t-\overline e_c\right] \\
&= \beta^T \Cov\left[\overline X_t-\overline X_c\right] \beta + \left(1-R^2\right) \VCR 
\end{align*}
We then bound the first term in the variance of $\widehat\tau$.

Recall that
\begin{align*}
    M = \frac{n}{n_tn_c} \left( W-\frac{n_t}{n}1_n \right)^T X S_{XX}^{-1} X^T \left( W-\frac{n_t}{n}1_n \right).
\end{align*}
We define 
\begin{align*}
    Z = \sqrt{\frac{n}{n_tn_c}} S_{XX}^{-1/2} X^T \left( W-\frac{n_t}{n}1_n \right) = \sqrt{\frac{n}{n_tn_c}} S_{XX}^{-1/2} (\overline X_t - \overline X_c).
\end{align*}
Then we have $M = Z^T Z$. 

$E[Z] = \sqrt{\frac{n}{n_tn_c}} S_{XX}^{-1/2} X^T E[W-\frac{1}{2}1_n] = 0$. Therefore, $E[Z_j] = 0$.

We note that the rows of $S_{XX}^{-1/2} X^T$ are exchangeable since the covariates are normalized. As a result, $\{Z_j\}_{1 \le j \le p}$ are exchangeable, and we have
\begin{align*}
    \Var[Z_j] = E[Z_j^2] = \frac{1}{p}E\left[\sum_{j = 1}^p Z_j^2\right] = \frac{1}{p}E[M] \le \frac{a}{p}.
\end{align*}
If we flip the sign of the $j$-th normalized covariate, the sign of $Z_j$ is also flipped, but the mapping $M(W)$ remains unchanged. Since the updating rule only depends on the Mahalanobis distance, this operation does not affect the distribution of $W$, hence the distribution of $Z_j$. As a result, $(Z_j,Z_l)$ and $(-Z_j,Z_l)$ are identically distributed. And we have
\begin{align*}
\Cov[Z_j, Z_l] &= E[Z_j Z_l] \\
&= E[E\left[Z_j Z_l \mid Z_l\right]] \\
&= E[Z_l E\left[Z_j \mid Z_l\right]] \\
&= E[Z_j \times 0] \\
&= 0 .
\end{align*}
Therefore, $\Cov[Z] = \Var[Z_j] I_p$, and
\begin{align*}
\beta^T \Cov[\overline X_t-\overline X_c] \beta &= \beta^T \Cov\left[ \sqrt{\frac{n_tn_c}{n}}S_{XX}^{1/2}Z \right] \beta \\
&= \frac{n_tn_c}{n} \beta^T S_{XX}^{1/2} \Cov[Z] S_{XX}^{1/2} \beta\\
&= \frac{n_tn_c}{n} \Var[Z_j] \beta^T S_{XX} \beta \\
&= \Var[Z_j] R^2 \VCR \\
&\le \frac{a}{p}R^2\VCR.
\end{align*}
Consequently,
\begin{align*}
    \Var[\widehat\tau] \le \frac{a}{p}R^2\VCR + (1-R^2)\VCR.
\end{align*}
Rearranging the terms, we obtain
\begin{align*}
    \frac{\VCR-\Var[\widehat\tau]}{\VCR} \ge \left( 1-\frac{a}{p} \right) R^2.
\end{align*}

\subsection{Proof of the Propositions}
Recall that $\mathcal{F}_0 = \{W \in \{0,1\}^n: \sum_i W_i = n_t\}$. Under Condition \ref{cond:equal sizes}, for any $W \in \mathcal{F}_0$,
\begin{align*}
    \sum_i (1-W_i) = n - n_t = n_c = n_t.
\end{align*}
This implies that $1-W \in \mathcal{F}_0$ as well. Thus, $\mathcal{F}_0 = 1 - \mathcal{F}_0$. Since CR, ARSRR, PSRR, and BRAIN treat $\mathcal{F}_0$ as the feasible set, they all satisfy Property \ref{prop:symmetric space}.

StratBRAIN and ClustBRAIN conduct rerandomization with more constraints imposed by the structure of the experiments. We show that these constraints are also symmetric in nature.

We denote the feasible set of StratBRAIN and ClustBRAIN as $\mathcal{F}_{Strat}$ and $\mathcal{F}_{Clust}$, respectively. 

Under Condition \ref{cond:equal sizes seq}, for any $W \in \mathcal{F}_{Strat}$,
\begin{align*}
    \sum_{i \in \mathcal{I}_k} (1-W_{i}) = n_k-n_{tk} = n_{ck} = n_{tk}.
\end{align*}
This implies that $1-W \in \mathcal{F}_{Strat}$ and $\mathcal{F}_{Strat} = 1 - \mathcal{F}_{Strat}$.

Under Condition \ref{cond:equal sizes clust}, for any $W \in \mathcal{F}_{Clust}$, suppose $U$ represents the assignment status of the clusters that corresponds to $W$. Then $1-U$ corresponds to $1-W$. And we have
\begin{align*}
    \sum_{k=1}^K (1-U_k) = K-K_t = K_c = K_t.
\end{align*}

This implies that $1-W \in \mathcal{F}_{Clust}$ and $\mathcal{F}_{Clust} = 1 - \mathcal{F}_{Clust}$.

Therefore, both StratBRAIN and ClustBRAIN satisfy Property \ref{prop:symmetric space}.

All of the methods satisfy Property \ref{prop:uniform initialization} since they all start with a random assignment by uniformly sampling from the feasible set.

Regarding the updating rule, CR automatically satisfies Property \ref{prop:symmetric update} since it does not update the treatment assignment.

For ARSRR, we have
\begin{align*}
    u_t = \frac{1}{|\mathcal{F}|.}
\end{align*}
For PSRR, we have
\begin{align*}
    u_t = \left(\frac{M(w^{(t-1)})}{M(w^{(t)})}\right)^\gamma.
\end{align*}
Therefore, both ARSRR and PSRR satisfy Property \ref{prop:symmetric update}.

For BRAIN and its variants, $(M(w^{(0)}),\ldots,M(w^{(t-1)}))$ determines whether $W^{(t)}$ is obtained through the local search step or the perturbation step.

In the local search step, we have
\begin{align*}
    u_t = 1\{M(w^{(t)})<M(w^{(t-1)})\}.
\end{align*}
And in the perturbation step, $u_t$ is a constant which depends on the number of adjacent feasible solutions.

Therefore, BRAIN satisfies Property \ref{prop:symmetric update}.

\subsection{Theoretical Results for SeqBRAIN}
We consider the following equal-group-size condition for sequentially randomized experiments. We note that it has the same form as the condition for stratified randomized experiments due to the similarity between the notations in the two settings.
\setcounter{condition}{3}
\begin{condition}
    $n_{tk} = n_{ck} = \frac{n_k}{2}$, $k = 1, \ldots, K$.
\end{condition}

We also have the corresponding version of Properties \ref{prop:symmetric space}-\ref{prop:symmetric update}, stated as follows.
\begin{property} \label{prop:symmetric space seq} 
    For any $k$,
    $\mathcal{F}_{[k]} = 1 - \mathcal{F}_{[k]}$.
\end{property}

\begin{property} \label{prop:uniform initialization seq} 
    For any $k$ and $w \in \mathcal{F}_{[k]}$,
    \begin{align*}
    P\left(W_{[k]}^{(0)} = w\right) = \frac{1}{|\mathcal{F}_{[k]}|}.
\end{align*}
\end{property}

\begin{property} \label{prop:symmetric update seq}
    There exists a sequence of functions $\{u_t\}_{t \ge 1}$, such that for any $k$, $t$ and $(w^{(0)},\ldots,w^{(t)})$ that may appear consecutively in the generated chain for the $k$-th stage,
    \begin{align*}
        P\left(W_{[k]}^{(t)} = w^{(t)} \mid W_{[k]}^{(0)} = w^{(0)}, \ldots, W_{[k]}^{(t-1)} = w^{(t-1)}\right) = u_t(M_{[k]}(w^{(0)}),\ldots,M_{[k]}(w^{(t)})),
    \end{align*}
\end{property}




Based on these sufficient conditions, we obtain that
\begin{theorem} \label{thm:unbiasedness seq}
    If $W$ is generated by a rerandomization algorithm satisfying Properties \ref{prop:symmetric space seq}-\ref{prop:symmetric update seq}, then $E[\widehat\tau] = \tau$.
\end{theorem}

\begin{theorem} \label{thm:variance seq}
    Under Conditions \ref{cond:linear relationship} and \ref{cond:normality}, if $W$ is generated by a rerandomization algorithm satisfying Properties \ref{prop:symmetric space seq}-\ref{prop:symmetric update seq}, then
\begin{align*}
    \frac{\VCR-\Var[\widehat\tau]}{\VCR} \ge \left( 1-\frac{a_K}{p} \right) R^2,
\end{align*}
\end{theorem}

We omit the proofs of Theorems \ref{thm:unbiasedness seq} and \ref{thm:variance seq}, since they are similar to the proofs of Theorems \ref{thm:unbiasedness} and \ref{thm:variance}, except that the chain $(W^{(0)},\ldots,W^{(T)})$ needs to be decomposed into $K$ parts corresponding to the $K$ stages. Each part can be analyzed in the same way as in simple randomized experiments.

We have verified that BRAIN satisfies Properties \ref{prop:symmetric space}-\ref{prop:symmetric update}. We note that SeqBRAIN can be viewed as a special version of BRAIN in each of its stages. As a result, SeqBRAIN satisfies the sequential version of Properties \ref{prop:symmetric space}-\ref{prop:symmetric update}.
\begin{proposition}
    SeqBRAIN satisfies Properties \ref{prop:uniform initialization seq} and \ref{prop:symmetric update seq}. Under Condition \ref{cond:equal sizes seq}, SeqBRAIN satisfies Property \ref{prop:symmetric space seq}.
\end{proposition}

The unbiasedness and variance reduction properties of SeqBRAIN are then immediately evident as corollaries of Theorems \ref{thm:unbiasedness seq} and \ref{thm:variance seq}.

\section{Randomization-based Inference} \label{app:inference}
We use the Fisher randomization test to test the sharp null hypothesis, $H_0: Y_i(1) - Y_i(0) = 0, i = 1, \ldots, n$, versus the two-sided alternative hypothesis, $H_1: Y_i(1) - Y_i(0) \neq 0, i = 1, \ldots, n$. As mentioned in Section \ref{sec:related}, the observed outcome for unit $i$ is $Y_i = W_iY_i(1) + (1-W_i)Y_i(0)$. Therefore, under $H_0$, $Y_i(1) = Y_i(0) = Y_i$, and we can impute all the potential outcomes with the observed outcomes. We then know the distribution of $\widehat\tau(W) = \frac{1}{n_t}\sum_{i:W_i=1}Y_i(1) - \frac{1}{n_c}\sum_{i:W_i=0}Y_i(0)$ since we can calculate $\widehat\tau(W)$ for every acceptable $W$. However, it is unrealistic to calculate the exact distribution of $\widehat\tau(W)$ due to the large size of the feasible set. As a result, we approximate the distribution of $\widehat\tau(W)$ using the Monte Carlo method. We sample $B=1000$ acceptable assignments $\{W^b\}_{1 \le b \le B}$ independently and calculate $\{\widehat\tau(W^b)\}_{1 \le b \le B}$. The p-value is then calculated by comparing $\{\widehat\tau(W^b)\}_{1 \le b \le B}$ with the observed treatment effect estimator $\widehat\tau(W^{obs})$, $pv = \frac{1}{B} \sum_{b=1}^B 1\{|\widehat\tau(W^b)| \ge |\widehat\tau(W^{obs})|\}$

As pointed out by \cite{zhu2022pair}, randomization-based confidence intervals can be constructed by inverting Fisher randomization tests. To construct a lower confidence bound for $\tau$, we consider testing $H_0:Y_i(1)-Y_i(0) = \theta, i = 1, \ldots, n$, versus $H_1:Y_i(1)-Y_i(0) > \theta, i = 1, \ldots, n$. Similar to testing the sharp null hypothesis with $\theta=0$, we can impute all the potential outcomes under $H_0$. We sample $B=1000$ acceptable assignments $\{W^b\}_{1 \le b \le B}$ independently and calculate the corresponding treatment effect estimators $\{\widehat\tau(W^b,\theta)\}_{1 \le b \le B}$ using the imputed potential outcomes. The p-value is then calculated as $pv(\theta) = \frac{1}{B} \sum_{b=1}^B 1\{|\widehat\tau(W^b,\theta)| \ge |\widehat\tau(W^{obs})|\}$. \cite{luo2021leveraging} proved that $\theta_l = \sup\{\theta: pv(\theta) \le \alpha\}$ is a lower confidence bound for $\tau$ with confidence level $\alpha$. And \cite{zhu2022pair} proposed an efficient approach to solve for $\theta_l$, which we adopt in our numerical experiments. Similarly, we can construct an upper confidence bound $\theta_u$ with confidence level $\alpha$. $[\theta_l, \theta_u]$ is then a confidence interval for $\tau$ with confidence level $1-2\alpha$.

\newpage
\section{Simulation Results} \label{app:results}
In \ref{app:results-complete}, we provide the complete results of the simulation studies in the main text. In \ref{app:results-randomness}, we examine the randomness of assignments generated by each method. In \ref{app:results-hyperparameter}, we compare different combinations of hyperparameters. In \ref{app:results-semi}, we generate semi-synthetic data based on the clinical trial dataset and compare the statistical performance of each method.

\subsection{Statistical and Computational Performances} \label{app:results-complete}
Table \ref{tab supp:simple} and Table \ref{tab supp:structured} show the complete results of our simulation studies for simple experiments and structured experiments, respectively. The results of ARSRR in the last setting of Table \ref{tab supp:simple} where $(n,p,p_a)=(500,250,10^{-4})$ are omitted due to the substantial computational burden. The overall findings are similar to the conclusion in the main text. BRAIN substantially reduces the computational burden of rerandomization and maintains the appealing statistical properties of rerandomization.

\subsection{Randomness Metric} \label{app:results-randomness}
Similar to \cite{zhu2022pair}, we denote $L_n$ as the largest eigenvalue of the covariance matrix of $2W-1$ and use $L_n$ to measure the randomness of assignments generated by each method. 
As pointed out by \cite{zhu2022pair}, larger $L_n$ implies less randomness.

Table \ref{tab supp:simple randomness} and Table \ref{tab supp:structured randomness} show the results in simple experiments and structured experiments, respectively. Similar to Table \ref{tab supp:simple}, the results of ARSRR in the last setting of Table \ref{tab supp:simple randomness} where $(n,p,p_a)=(500,250,10^{-4})$ are omitted due to the substantial computational burden. CR is the most random method as expected. The three rerandomization methods perform similarly in all settings. Therefore, one can safely perform randomization-based statistical inference based on the assignments generated by BRAIN without the risk of relying on a small set of unique assignments.

\subsection{Hyperparameters} \label{app:results-hyperparameter}
There are $n$ units, hence we can swap at most $n/2$ disjoint pairs in an iteration. So, the maximum value of $L$ and $S$ is $n/2$. The minimum value of $L$ and $S$ is 1. We also consider the maximum value times 0.25, 0.5, and 0.75. Overall, there are $5 \times 5 = 25$ combinations in total.

\begin{figure}[htbp]
    \centering
    \includegraphics[width=0.9\linewidth]{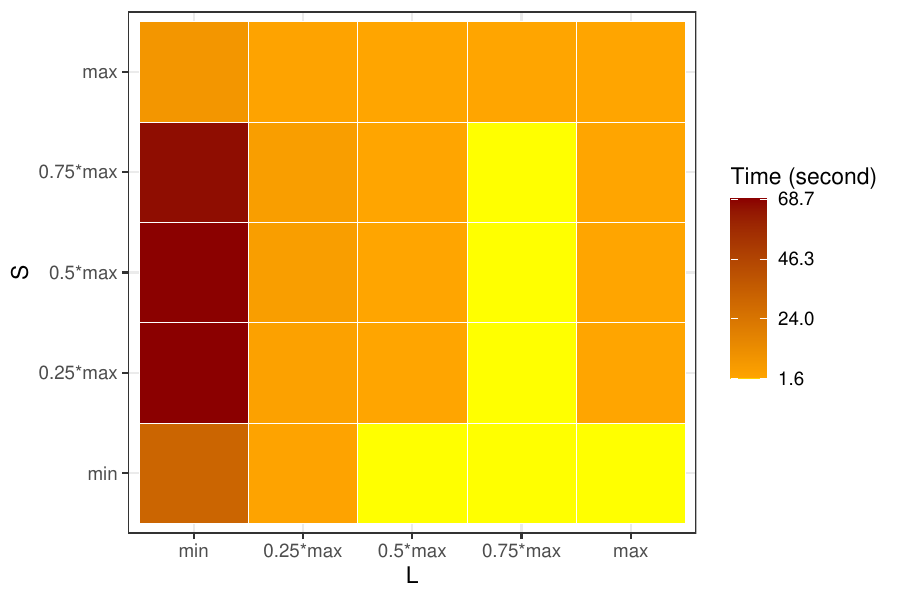}
    \caption{Comparison of different hyperparameters.}
    \label{fig:hyperparameter}
\end{figure}

Figure \ref{fig:hyperparameter} shows the computational time of BRAIN with different combinations of $L$ and $S$ under the setting where $(n,p,p_a) = (30,15,10^{-3})$. It demonstrates that our choice of $(L,S)$ is among the top tier.

\subsection{Semi-synthetic Data} \label{app:results-semi}
We mimic the set-ups in \cite{zhu2022pair} and simulate semi-synthetic data based on the clinical trial dataset. Since we only observe half of the potential outcomes in the original dataset, we impute the counterfactual potential outcomes under the sharp null. We then compare BRAIN and the existing methods in the same way as in our simulation studies. The results are shown in Table \ref{tab supp:semi} and are similar to our findings in the simulation studies. BRAIN significantly accelerates rerandomization and has comparable performances in other aspects compared to ARSRR and PSRR.

\newpage
\begin{table}[htbp]
\caption{Statistical and computational performance of different methods in simple randomized experiments. SD: standard deviation. CP: coverage probability. Len: length.} \label{tab supp:simple}
\begin{center}
\begin{tabular}{ccclccccccccc}
\toprule
         \multirow{2}{*}{$n$} & \multirow{2}{*}{$p$} & \multirow{2}{*}{$p_a$} & \multirow{2}{*}{Method} & & \multicolumn{6}{c}{Statistical Inference ($\times 10^{-3}$)} & & \multirow{2}{1.7cm}{\parbox{1.7cm}{\vspace{1ex}\centering Run Time \\ \centering (second)}} \\
         \cmidrule{6-11}
          & & & & & Bias & SD & Size & Power & CP & Len & & \\
          \hline
         \multirow{8}{*}{30} & \multirow{4}{*}{2} & \multirow{4}{*}{$10^{-3}$} & CR & & 6.2 & 74 & 10.5 & 14 & 89.5 & 246 & & $<0.1$ \\
         & & & ARSRR & & 0.1 & 51 & 8.7 & 19 & 91.3 & 177 & & 11.4 \\
         & & & PSRR & & 1.5 & 51 & 10.1 & 18 & 89.9 & 177 & & 48.1 \\
         & & & BRAIN & & 1.6 & 52 & 11.0 & 17 & 89.0 & 177 & & \textbf{1.4} \\
         \cmidrule{2-13}
         & \multirow{4}{*}{15} & \multirow{4}{*}{$10^{-3}$} & CR & & 9.2 & 199 & 9.3 & 18 & 90.7 & 678 & & $<0.1$ \\
         & & & ARSRR & & 2.2 & 155 & 10.8 & 21 & 89.2 & 552 & & 64.2 \\
         & & & PSRR & & 3.8 & 156 & 10.6 & 22 & 89.4 & 547 & & 1.6 \\
         & & & BRAIN & & 6.8 & 152 & 10.4 & 22 & 89.6 & 547 & & \textbf{0.3} \\
         \hline
         \multirow{16}{*}{100} & \multirow{8}{*}{2} & \multirow{4}{*}{$10^{-3}$} & CR & & 1.4 & 42 & 11.7 & 29 & 88.3 & 132 & & $<0.1$ \\
         & & & ARSRR & & 0.2 & 29 & 10.4 & 40 & 89.6 & 94 & & 14.0 \\
         & & & PSRR & & 0.3 & 29 & 9.6 & 40 & 90.4 & 94 & & 2.1 \\
         & & & BRAIN & & 0.4 & 27 & 8.5 & 40 & 91.5 & 94 & & \textbf{0.5} \\
         \cmidrule{3-13}
         & & \multirow{4}{*}{$10^{-4}$} & CR & & 1.4 & 42 & 11.7 & 29 & 88.3 & 132 & & $<0.1$ \\
         & & & ARSRR & & 0.0 & 28 & 9.8 & 40 & 90.2 & 94 & & 139.6 \\
         & & & PSRR & & 0.1 & 28 & 10.1 & 40 & 89.9 & 94 & & 35.7 \\
         & & & BRAIN & & 0.3 & 28 & 10.2 & 41 & 89.8 & 94 & & \textbf{4.2} \\
         \cmidrule{2-13}
         & \multirow{8}{*}{50} & \multirow{4}{*}{$10^{-3}$} & CR & & 1.0 & 207 & 11.3 & 35 & 88.7 & 663 & & $<0.1$ \\
         & & & ARSRR & & 1.4 & 174 & 9.9 & 39 & 90.1 & 579 & & 215.3 \\
         & & & PSRR & & 5.1 & 170 & 10.1 & 42 & 89.9 & 574 & & 8.2 \\
         & & & BRAIN & & 2.1 & 173 & 10.1 & 40 & 89.9 & 573 & & \textbf{0.2} \\
         \cmidrule{3-13}
         & & \multirow{4}{*}{$10^{-4}$} & CR & & 1.0 & 207 & 11.3 & 35 & 88.7 & 663 & & $<0.1$ \\
         & & & ARSRR & & 10.2 & 165 & 7.8 & 39 & 92.2 & 567 & & 2453.3 \\
         & & & PSRR & & 2.8 & 167 & 9.9 & 43 & 90.1 & 560 & & 27.0 \\
         & & & BRAIN & & 6.4 & 163 & 9.1 & 42 & 90.9 & 558 & & \textbf{0.3} \\
         \hline
         \multirow{16}{*}{500} & \multirow{8}{*}{2} & \multirow{4}{*}{$10^{-3}$} & CR & & 1.0 & 18 & 8.7 & 74 & 91.3 & 59 & & $<0.1$ \\
         & & & ARSRR & & 0.4 & 13 & 8.8 & 94 & 91.2 & 42 & & 24.3 \\
         & & & PSRR & & 0.6 & 12 & 9.1 & 94 & 90.9 & 42 & & 1.6 \\
         & & & BRAIN & & 0.1 & 13 & 10.5 & 93 & 89.5 & 42 & & \textbf{0.5} \\
         \cmidrule{3-13}
         & & \multirow{4}{*}{$10^{-4}$} & CR & & 1.0 & 18 & 8.7 & 74 & 91.3 & 59 & & $<0.1$ \\
         & & & ARSRR & & 0.5 & 12 & 0.2 & 95 & 90.8 & 42 & & 248.8 \\
         & & & PSRR & & 0.5 & 13 & 11.2 & 92 & 88.8 & 42 & & 6.4 \\
         & & & BRAIN & & 0.2 & 12 & 11.3 & 92 & 88.7 & 42 & & \textbf{1.7} \\
         \cmidrule{2-13}
         & \multirow{8}{*}{250} & \multirow{4}{*}{$10^{-3}$} & CR & & 0.8 & 201 & 10.4 & 73 & 89.6 & 660 & & $<0.1$ \\
         & & & ARSRR & & 4.0 & 186 & 9.2 & 77 & 90.8 & 619 & & 4137.4 \\
         & & & PSRR & & 4.5 & 189 & 9.8 & 78 & 90.2 & 619 & & 275.3 \\
         & & & BRAIN & & 7.6 & 199 & 12.6 & 73 & 87.4 & 615 & & \textbf{1.2} \\
         \cmidrule{3-13}
         & & \multirow{3}{*}{$10^{-4}$} & CR & & 0.8 & 201 & 10.4 & 88 & 89.6 & 660 & & $<0.1$ \\
         & & & PSRR & & 4.3 & 187 & 10.3 & 91 & 89.7 & 611 & & 2878.8 \\
         & & & BRAIN & & 5.7 & 181 & 9.8 & 92 & 90.2 & 608 & & \textbf{1.3} \\
    \bottomrule
\end{tabular}
\end{center}
\end{table}

\newpage

\begin{table}[htbp]
\caption{Statistical and computational performance of different methods in structured randomized experiments. SD: standard deviation. CP: coverage probability. Len: length.} \label{tab supp:structured}
\begin{center}
\begin{tabular}{cclccccccccc}
\toprule
         \multirow{2}{*}{Scheme} & \multirow{2}{*}{$n$} & \multirow{2}{*}{Method} & & \multicolumn{6}{c}{Statistical Inference ($\times 10^{-3}$)} & & \multirow{2}{1.7cm}{\parbox{1.7cm}{\vspace{1ex}\centering Run Time \\ \centering (second)}} \\
         \cmidrule{5-10}
          & & & & Bias & SD & Size & Power & CP & Len & & \\
          \hline
         \multirow{8}{*}{Sequential} 
         & \multirow{4}{*}{200} & CR & & 5.5 & 140 & 10.3 & 28 & 89.7 & 469 & & $<0.1$ \\
         & & ARSRR & & 3.3 & 112 & 8.1 & 36 & 91.9 & 391 & & 162.4 \\
         & & PSRR & & 1.7 & 118 & 10.4 & 36 & 89.6 & 391 & & 14.1 \\
         & & BRAIN & & 0.9 & 116 & 10.2 & 35 & 89.8 & 390 & & \textbf{0.9} \\
         \cmidrule{2-12}
         & \multirow{4}{*}{1000} & CR & & 3.2 & 146 & 10.7 & 76 & 89.3 & 466 & & $<0.1$ \\
         & & ARSRR & & 6.0 & 139 & 12.2 & 80 & 87.8 & 428 & & 3288.5 \\
         & & PSRR & & 3.3 & 135 & 11.5 & 82 & 88.5 & 429 & & 339.1 \\
         & & BRAIN & & 5.8 & 138 & 11.4 & 81 & 88.6 & 428 & & \textbf{5.8} \\
         \hline
         \multirow{6}{*}{Stratified} 
         & \multirow{3}{*}{200} & CR & & 5.2 & 140 & 10.3 & 26 & 89.7 & 467 & & $<0.1$ \\
         & & ARSRR & & 4.5 & 121 & 10.6 & 35 & 89.4 & 401 & & 148.0 \\
         & & BRAIN & & 6.1 & 124 & 10.3 & 32 & 89.7 & 402 & & \textbf{0.5} \\
         \cmidrule{2-12}
         & \multirow{3}{*}{1000} & CR & & 3.0 & 140 & 10.0 & 76 & 90.0 & 466 & & $<0.1$ \\
         & & ARSRR & & 2.7 & 131 & 9.1 & 82 & 90.9 & 434 & & 3386.8 \\
         & & BRAIN & & 0.0 & 130 & 10.5 & 83 & 89.5 & 436 & & \textbf{3.0} \\
         \hline
         \multirow{6}{*}{Cluster} 
         & \multirow{3}{*}{200} & CR & & 3.0 & 141 & 10.9 & 27 & 89.1 & 479 & & $<0.1$ \\
         & & ARSRR & & 3.3 & 116 & 8.8 & 32 & 91.2 & 405 & & 126.4 \\
         & & BRAIN & & 6.4 & 116 & 8.9 & 30 & 91.1 & 405 & & \textbf{0.8} \\
         \cmidrule{2-12}
         & \multirow{3}{*}{1000} & CR & & 0.9 & 140 & 9.8 & 77 & 90.2 & 465 & & $<0.1$ \\
         & & ARSRR & & 6.2 & 132 & 9.3 & 80 & 90.7 & 435 & & 3116.2 \\
         & & BRAIN & & 3.3 & 131 & 10.1 & 82 & 89.9 & 436 & & \textbf{3.9} \\
    \bottomrule
\end{tabular}
\end{center}
\end{table}

\begin{table}[htbp]
\caption{$L_n$ ($\times 10^{-2}$) of different methods in simple randomized experiments.} \label{tab supp:simple randomness}
\begin{center}
\begin{tabular}{cccccccc}
\toprule
         \multirow{2}{*}{$n$} & \multirow{2}{*}{$p$} & \multirow{2}{*}{$p_a$} & & \multicolumn{4}{c}{Method} \\
         \cmidrule{5-8}
          & & & & CR & ARSRR & PSRR & BRAIN \\
          \hline
         \multirow{2}{*}{30} & 2 & $10^{-3}$ & & 137 & 158 & 161 & 158 \\
         \cmidrule{2-8}
         & 15 & $10^{-3}$ & & 137 & 347 & 333 & 329 \\
         \hline
         \multirow{4}{*}{100} & \multirow{2}{*}{2} & $10^{-3}$ & & 170 & 174 & 174 & 174 \\
         \cmidrule{3-8}
         & & $10^{-4}$ & & 170 & 174 & 174 & 174 \\
         \cmidrule{2-8}
         & \multirow{2}{*}{50} & $10^{-3}$ & & 170 & 231 & 234 & 233 \\
         \cmidrule{3-8}
         & & $10^{-4}$ & & 170 & 242 & 248 & 245 \\
         \hline
         \multirow{4}{*}{500} & \multirow{2}{*}{2} & $10^{-3}$ & & 288 & 289 & 289 & 289 \\
         \cmidrule{3-8}
         & & $10^{-4}$ & & 288 & 289 & 289 & 289 \\
         \cmidrule{2-8}
         & \multirow{2}{*}{250} & $10^{-3}$ & & 288 & 307 & 307 & 310 \\
         \cmidrule{3-8}
         & & $10^{-4}$ & & 288 & NA & 314 & 317 \\
    \bottomrule
\end{tabular}
\end{center}
\end{table}

\begin{table}[htbp]
\caption{$L_n$ ($\times 10^{-2}$) of different methods in structured randomized experiments.} \label{tab supp:structured randomness}
\begin{center}
\begin{tabular}{ccccccc}
\toprule
         \multirow{2}{*}{Scheme} & \multirow{2}{*}{$n$} & & \multicolumn{4}{c}{Method} \\
         \cmidrule{4-7}
          & & & CR & ARSRR & PSRR & BRAIN \\
          \hline
         \multirow{2}{*}{Sequential} 
         & 200 & & 208 & 251 & 252 & 253 \\
         \cmidrule{2-7}
         & 1000 & & 396 & 411 & 410 & 412\\
         \hline
         \multirow{2}{*}{Stratified} 
         & 200 & & 208 & 228 & NA & 228 \\
         \cmidrule{2-7}
         & 1000 & & 396 & 404 & NA & 404 \\
         \hline
         \multirow{2}{*}{Cluster} 
         & 200 & & 339 & 438 & NA & 430 \\
         \cmidrule{2-7}
         & 1000 & & 576 & 606 & NA & 604 \\
    \bottomrule
\end{tabular}
\end{center}
\end{table}

\begin{table}[htbp]
\caption{Statistical and computational performance of different methods when applied to semi-synthetic data. SD: standard deviation. CP: coverage probability. Len: length.} \label{tab supp:semi}
\begin{center}
\begin{tabular}{lccccccccc}
\toprule
         \multirow{2}{*}{Method} & & \multicolumn{5}{c}{Statistical Inference ($\times 10^{-3}$)} & & \multirow{2}{2.4cm}{Computation Time (second)} & \multirow{2}{*}{$L_n$} \\
         \cmidrule{3-7}
          & & Bias & SD & Size & CP & Len & & \\
          \hline
         CR & & 9.7 & 440 & 4.4 & 95.6 & 1804 & & $<0.1$ & 122 \\
         ARSRR & & 2.9 & 317 & 4.3 & 95.7 & 1333 & & 13.9 & 140 \\
         PSRR & & 0.6 & 319 & 4.8 & 95.2 & 1346 & & 219.9 & 142 \\
         BRAIN & & 14.4 & 336 & 5.3 & 94.7 & 1338 & & \textbf{1.2} & 139 \\
    \bottomrule
\end{tabular}
\end{center}
\end{table}

\end{document}